\documentclass[prl,twocolumn,showpacs,showkeys,secnumarabic,amssymb,amsmath,superscriptaddress,floatfix]{revtex4-1}
\usepackage{amssymb,amsmath}
\usepackage{color}
\usepackage{tikz}
\usepackage{mathtools}
\usetikzlibrary{arrows.meta}
\usepackage{graphics}
\usepackage{graphicx}
\usepackage{enumitem}
\usepackage{bm}
\usepackage{standalone}
\usepackage[toc,page]{appendix}

\begin{document}

\title{Characterizing coherence with quantum observables}

\author{Suman Mandal}
\affiliation{Division of Arts and Sciences, NYU Shanghai, 1555 Century Ave, Shanghai, 200122, China.}

\author{Marek Narozniak} 
\affiliation{Division of Arts and Sciences, NYU Shanghai, 1555 Century Ave, Shanghai, 200122, China.}
\affiliation{Department of Physics, New York University, New York, NY, 10003, USA.}

\author{Chandrashekar Radhakrishnan}
\affiliation{Division of Arts and Sciences, NYU Shanghai, 1555 Century Ave, Shanghai, 200122, China.}

\author{Zhiqiang Jiao}
\affiliation{State Key Laboratory of Advanced Optical
Communication Systems and Networks, School of Physics and Astronomy,
Shanghai Jiao Tong University, Shanghai 200240, China.}

\author{Xianmin Jin}
\affiliation{State Key Laboratory of Advanced Optical
Communication Systems and Networks, School of Physics and Astronomy,
Shanghai Jiao Tong University, Shanghai 200240, China.}

\author{Tim Byrnes}
\email{tim.byrnes@nyu.edu}
\affiliation{Division of Arts and Sciences, NYU Shanghai, 1555 Century Ave, Shanghai, 200122, China.}
\affiliation{Department of Physics, New York University, New York, NY, 10003, USA.}

\begin{abstract}
We introduce a procedure based on quantum expectation values of measurement observables to characterize quantum coherence.   Our measure allows one to quantify coherence without having to perform tomography of the quantum state, and can be directly calculated from measurement expectation values.  
This definition of coherence allows the decomposition into contributions corresponding to the non-classical correlations between the subsystems and localized on each subsystem.  The method can also be applied to cases where the full set of measurement operators is unavailable. An estimator using the truncated measurement operators can be used to obtain lower bound to the genuine value of coherence.
We illustrate the method for several bipartite systems, and show the singular behavior of the coherence measure in a spin-1 chain, characteristic of a quantum phase transition. 
\end{abstract}


\date{\today}

\maketitle

\paragraph{Introduction}

Coherence is one of the fundamental concepts of quantum mechanics and has been studied extensively in the context of 
phase space distributions \cite{glauber1963coherent} and correlation functions \cite{sudarshan1963equivalence,scully1999quantum}.  Despite its long history, it was not quantified in a formal sense until recently using the tools of quantum information theory \cite{baumgratz2014quantifying}.  These ideas have led to many new developments regarding quantum measurement \cite{shao2015fidelity,rana2016trace,napoli2016robustness,girolami2014observable}, the distribution of coherence in multipartite systems \cite{radhakrishnan2016distribution}, and its application for characterizing states  \cite{zheng2016detecting,mani2015cohering,streltsov2016entanglement,malvezzi2016quantum,
radhakrishnan2017quantum,radhakrishnan2017quantum2,zhang2018estimating,carmeli2018determining,lami2018generic,lami2019completing}.  In particular, there has been a lot of attention to the development of the resource theory of coherence 
\cite{winter2016operational,yadin2016quantum,chitambar2016critical,streltsov2017colloquium,de2016genuine,aberg2006quantifying}.

One of the promising applications of the theory of quantum coherence is in characterizing the nature of complex many-body systems.  While many variations exist for suitable measures to quantify coherence, such as the $ l_1 $-norm, relative entropy \cite{baumgratz2014quantifying}, and the Jensen-Shannon divergence \cite{radhakrishnan2016distribution}, all methods require complete knowledge of the density matrix of the system.   While this is not difficult 
for systems involving a few qubits, the computational cost of calculating coherence in systems with large 
Hilbert spaces becomes prohibitive to make it a useful tool for characterizing the coherence.  The situation is made worse when it is desirable to find the distribution of coherence, finding the contributions that lie on local subsystems and collectively between them.  Evaluating these contributions can involve evaluating the minimum of complex optimization problems \cite{radhakrishnan2016distribution}, which make them difficult to use in a practical sense.  

In this paper, we introduce a method of quantifying the quantum coherence and its distribution in a bipartite system
using expectation values of physical observables.   The primary advantage of this approach is that it does not explicitly require tomographic reconstruction of a density matrix, which may be  difficult or computationally expensive.  The expectation values of a set of observables can be directly used to calculate the coherence.   We also show that it is possible to estimate the coherence in high dimensional systems, by considering a truncated set of measurement observables.  In bipartite systems, by choosing the observables to be the form of a correlation function, it allows one to find the contribution of the coherence due to non-classical correlations and the amount localized on each subsystem \cite{radhakrishnan2016distribution,tan2016unified,tan2018entanglement,ma2017accessible,
kraft2018genuine,radhakrishnan2017quantum,radhakrishnan2018basis,bu2017distribution,radhakrishnan2019time}.  The contribution due to correlations is found to be of a form that is the difference of two 
covariance matrices \cite{duan2000inseparability,simon2000peres,braunstein2005quantum,wang2007quantum,guhne2007covariance,gittsovich2008unifying,toth2010separability,gittsovich2010quantifying}.  The same set of measurements that are used to construct the coherence measure can also be used to construct the covariance matrix, which has been used as an effective way of detecting entanglement \cite{duan2000inseparability,simon2000peres,braunstein2005quantum,wang2007quantum,guhne2007covariance,gittsovich2008unifying,toth2010separability,gittsovich2010quantifying,guhne2009entanglement,vinay2018}.  Our approach in characterizing coherence with observables extends the toolbox for characterizing the features of a quantum state.

\paragraph{Observable based measure of coherence}

Consider an arbitrary quantum state $ \rho $ in a $ D $-dimensional Hilbert space with an orthonormal set of basis vectors $ \{ |k \rangle : k \in [1,D]  \}$.  We start by writing the the density matrix in terms of expectation values of a set of orthonormal observables.
\begin{align}
\Omega_{k k'} (\rho) = \frac{e^{i \pi/4} \langle {\cal M}_{k k'} \rangle_\rho 
+ e^{-i \pi/4} \langle {\cal M}_{k' k} \rangle_\rho }{\sqrt{2}e^{i \pi \text{sgn}(k'-k)/4}} ,
\label{omegadef}
\end{align}
where $ {\cal M}_{k k'} $ are a set of observables which form an orthonormal operator basis $ \text{Tr} ( {\cal M}_{j j'} {\cal M}_{k k'} ) = \delta_{j j'} \delta_{k k'} $.  A standard choice of 
observable operators is \cite{gittsovich2010quantifying}
\begin{align}
{\cal M}_{k k'} = \left\{
\begin{array}{cc}
| k \rangle \langle k | & \hspace{5mm} k'=k \\
 ( |k \rangle \langle k' | + |k'\rangle \langle k |)/\sqrt{2} & \hspace{5mm} k'> k  \\
 ( i|k \rangle \langle k' | -i|k' \rangle \langle k |)/\sqrt{2} & \hspace{5mm} k'< k 
\end{array}
\right.  .
\label{standardmeasurementops}
\end{align}
 One can verify that the observable matrix coincides with the original density matrix $ \Omega (\rho) = \rho $ for the case that the measurement operators are taken to be (\ref{standardmeasurementops}).   

To measure the quantum coherence in the system we must compare the state $ \rho $ to its decohered version $ \rho_d $, which is obtained by setting of all off-diagonal terms to zero in a chosen basis. Using the observable matrix, we can construct a coherence measure according to
\begin{align}
\label{covariancebasedmeasure}
C (\rho) = \| \Omega(\rho) - \Omega(\rho^{d}) \|_1 ,
\end{align}
where $ \| \cdot \|_1 $ is the Schatten $1$-norm (or trace norm). We opt for the Schatten-$1$-norm since it is an invariant quantity under unitary transformations of the state $ C(\rho) = C( U^\dagger \rho U) $ and orthogonal transformations over the standard operators ${\cal M}_{kk'} $ (see Supplementary Material).   We note this is not the same as recent discussions regarding basis independent coherence \cite{radhakrishnan2019basis,ma2019operational,wang2017intrinsic}. Our measure is explicitly basis dependent, defined by the decohered matrix $ \rho_d $.  The invariance of the Schatten $1$-norm will allow us to use an arbitrary set of measurement observables.  

Showing that (\ref{covariancebasedmeasure}) is a valid coherence measure can be straightforwardly proved using the alternative framework of Ref. \cite{yu2016alternative}.  The criteria proposed by Yu, Tong, and co-workers state that a valid coherence measure must satisfy: (C1$'$) $C (\rho)  \geq 0$ and $C(\rho) = 0$ iff $\rho \in \mathcal{I} $, where $ \mathcal{I} $ is the set of all incoherent states; (C2$'$) $C (\rho) \ge C(  \Lambda(\rho)) $ where $ \Lambda $ is an incoherent operation; (C3$'$) $ C( p_1 \rho_1 \oplus p_2 \rho_2 ) = p_1 C(\rho_1) + p_2  C(\rho_2) $ where $ \rho_1, \rho_2 $ are in independent subspaces and $ p_1 + p_2 = 1 $ are probabilities. These have been shown to be equivalent to the conditions in the original work of Ref. \cite{baumgratz2014quantifying}.  
We show in the Supplementary Material that all the conditions are satisfied for our coherence measure.    In short, (C1$'$) is satisfied due to the uniqueness of the observables $ \Omega(\rho) $ for a given quantum state,  (C2$'$) follows from the contractivity of the Schatten-$1$ norm \cite{perez2006contractivity}, (C3$'$) follows from the fact that the the Schatten-$1$ norm of a block diagonal matrix is simply the sum of the norms of the block diagonals.  As a further check, we have verified that the original coherence measure conditions of Ref. \cite{baumgratz2014quantifying} are satisfied.

The form of the coherence (\ref{covariancebasedmeasure}) is still not in a convenient form since it uses a specific choice of measurement operators $ {\cal M}_{k k'} $, which may not coincide with what is available (from an experiment, for example). Suppose the set of $ N $ available observables (Hermitian operators) are ${\cal O} = \{ {\cal S}_{l}: l \in[1,N] \}$ and are orthonormal $ \text{Tr} ( {\cal S}_l  {\cal S}_{l'} ) =  \delta_{ll'} $.  We first assume that the number of operators is complete, such that $ N = D^2 $.  Using the fact that one can expand the standard operators $ {\cal M}_{k k'} $ in terms of the operators $ {\cal S}_l $ according to  $  {\cal M}_{k k'} = \sum_l V_{k k'}^{(l)} {\cal S}_l $, 
where $ V_{k k'}^{(l)}  = \text{Tr} ( {\cal S}_l {\cal M}_{k k'} ) $, we can rewrite the observable matrix elements as 
\begin{align}
\Omega_{k k'} (\rho) & =  \sum_{l\in {\cal O}} \left( \frac{ e^{i \pi/4}  V_{k k'}^{(l)}  + e^{-i \pi/4}  V_{k' k}^{(l)} }{\sqrt{2} e^{i \pi \text{sgn}(k'-k)/4}} \right)  \langle {\cal S}_l \rangle_\rho  .
\end{align}
Substituting the definitions of $ V_{k k'}^{(l)} $ we arrive at the expression
\begin{align}
C (\rho) = \| \sum_{l\in {\cal O}} {\cal S}_l (   \langle {\cal S}_l \rangle_\rho - \langle {\cal S}_l \rangle_{\rho^d} ) \|_1  . 
\label{observables}
\end{align}
 This is the main expression for the coherence that we will use.  Thus the matrix to be evaluated is a linear combination of the measurement operators themselves, weighted by 
the expectation values of the state to be measured and its decohered counterpart.  All these are directly accessible from experiment, hence (\ref{observables}) constitutes a convenient way of calculating the coherence.

\paragraph{Decomposing coherence}

Now consider that the total system can be subdivided into two parts, which we label by $ A $ and $ B $, with dimension $ D_A $ and $ D_B $ respectively.  As a particular choice of measurement observables, let us take $ \{ {\cal S}_{ll'}  = {\cal A}_{l} \otimes  {\cal B}_{l'} : l \in [1,N_A], l'\in [1,N_B] \}$, where $  {\cal A}_{l},  {\cal B}_{l'} $ are operators on $ A, B $ respectively.  The total number of operators is $ N= N_A N_B $, where $ N = D_{A}^2 D_{B}^2 $ for a complete operator basis. The total coherence in this case is calculated in the same way as (\ref{covariancebasedmeasure}) with (\ref{observables}) summing over all indices $ l, l' $.  

Now consider the product state
\begin{align}
\pi_\rho \equiv \rho_A \otimes \rho_B,
\end{align}
which is obtained by finding the tensor product of the reduced density matrices.  Since this state has no correlations at all between $A$ and $ B$, any coherence that is present must be entirely due to local contributions.  We thus define the local coherence \cite{radhakrishnan2018basis} to be 
\begin{align}
C_{\text{L}} (\rho) \equiv C( \pi_\rho) & = \|\Omega( \pi_\rho) - \Omega( \pi_{\rho}^d) \|_1 \nonumber \\
& = \| \sum_{l l'} {\cal S}_{l l'} ( \langle {\cal S}_{l l'} \rangle_{\pi_\rho} - 
\langle {\cal S}_{l l'} \rangle_{\pi_\rho^d} ) \|_1 
\label{localcoherence}
\end{align}
where $ \pi_{\rho}^d = \rho_A^d \otimes \rho_B^d = \pi_{\rho^d } $ and $ \rho_{A,B}^d $ are the decohered reduced density matrices. 

We would now like to remove this contribution from the total coherence, hence we propose the quantity
\begin{align}
\delta(\rho) = \| \Omega(\rho) - \Omega(\rho^{d}) -  \Omega( \pi_\rho) + \Omega( \pi_{\rho}^d) \|_1 .
\label{quantumcorrelations}
\end{align}
This quantity can be written in an illuminating way by defining the covariance matrix \cite{duan2000inseparability,simon2000peres,braunstein2005quantum,wang2007quantum,guhne2007covariance,gittsovich2008unifying,toth2010separability,gittsovich2010quantifying,guhne2009entanglement,vinay2018}
\begin{align}
\gamma_{l l'}(\rho) = \langle {\cal A}_{l}  \otimes {\cal B}_{l'}  \rangle_\rho - \langle {\cal A}_{l}  \rangle_\rho \langle {\cal B}_{l'} \rangle_\rho  .
\end{align}
Eq. (\ref{quantumcorrelations}) can then be written
\begin{align}
\delta(\rho) = 
\| \sum_{l l'} {\cal S}_{l l'}  (\gamma_{l l'}(\rho) - \gamma_{l l'}(\rho^d)) \|_1 ,
\label{quantumcorrelations2}
\end{align}
which is the difference between the original and decohered covariance matrices.
 A covariance matrix characterizes the correlations between 
the two subsystems, including both quantum and classical contributions. The decohered state $ \rho^d $ contains no off-diagonal terms, hence is a completely classical state in terms of the correlations with respect to the basis choice of the decohered one. This means that $ \gamma(\rho^d) $ only contains classical correlations, while $ \gamma(\rho) $ contains both quantum and classical parts.  We can therefore interpret (\ref{quantumcorrelations2}) as a quantity which is related to the quantum correlations between the subsystems $ A $ and $ B $, and hence, we name $\delta(\rho)$ the ``global correlations''. It should be emphasized that the global correlations here are basis-dependent in the sense that the classical correlations {\it in a particular basis} are being subtracted.  Since the basis choice is fixed and not optimized such as in quantum discord, we do not expect to get equivalent results using (\ref{quantumcorrelations2}) alone \cite{ollivier2001quantum}.  

It is tempting to define (\ref{quantumcorrelations2}) as a type of coherence as it has been done in numerous past works \cite{radhakrishnan2016distribution,zheng2016detecting,mani2015cohering,streltsov2016entanglement,malvezzi2016quantum,radhakrishnan2017quantum,radhakrishnan2017quantum2}.  However, $ \delta(\rho) $ does not properly satisfy the coherence properties (see Supplementary Materials), hence it is not strictly appropriate to call it a type of coherence. The local coherence (\ref{localcoherence}) on the other hand has the same form as (\ref{covariancebasedmeasure}), hence is a coherence measure.  Using the triangle inequality of any matrix norm, the three quantities can be related as
\begin{align}
C(\rho) \le C_{\text{L}}(\rho) + \delta(\rho) .
\label{decomposition}
\end{align}
This relation allows us to decompose the coherence into its local contribution and parts originating from the global correlations between the subsystems.  

In certain limiting cases the decomposition (\ref{decomposition}) is guaranteed to give an equality relation. For example, for a product state possessing no correlations $ \rho = \rho_A \otimes \rho_B $, one can easily verify that $ \delta(\rho) = 0 $ and $ C(\rho) = C_{\text{L}}(\rho) $.  In the opposite limit of a maximally entangled bipartite state $ \rho = | \Psi_E \rangle \langle \Psi_E | $ with $ | \Psi_E \rangle = \sum_m |m \rangle  | m \rangle/\sqrt{D_A} $, where $ D_A = D_B $, the reduced density matrices are $ \rho_{A,B} = I/D_A $, and are diagonal. Hence, the local coherence $ C_{\text{L}}(\rho) = 0 $ and total coherence $ C(\rho) = \delta(\rho) $.

\paragraph{Incomplete sets of operators}

Up to the this point we have assumed that the set of operators $ \{ {\cal S}_l  \}$ forms a full operator basis.  An important question is what the effect of relaxing this requirement is, such that the number of observables does not have the complete information contained in the density matrix $ N < D^2 $.  This is relevant in the context of high dimensional systems, where it is difficult to perform full tomography of the quantum system.  When such a truncation is performed, it is natural to expect that not all the coherence in the system is captured by the coherence measure, since certain measurement operators that characterize the coherence may be missing.  Nevertheless, we demand that the truncation should be performed in a controlled way, such that the estimated value of the coherence does not have a spurious dependence on the truncation.  Specifically we demand that the truncated coherence satisfies $ C^{\text{tr}} (\rho) \le  C (\rho) $, such that the estimated value never overestimates the genuine value of coherence.  Unfortunately, simply truncating the number of operators in (\ref{observables}) violates this inequality due to the properties of Schatten-$1$ norm  \footnote{We note that instances of the violation are rather rare and the magnitude of the violation is small.  Hence if one is satisfied with an approximate lower bound, the Schatten-$1$ norm may be suitable in some cases. }.  However, using the Frobenius norm for the truncated measure satisfies the inequality, hence we define
\begin{align}
C^{\text{tr}} (\rho) = \| \sum_{l\in {\cal O}_{\text{tr}}} {\cal S}_l (   \langle {\cal S}_l \rangle_\rho - \langle {\cal S}_l \rangle_{\rho^d} ) \|_2
\label{truncatedc}
\end{align}
where $ {\cal O}_{\text{tr}} $ is the set of truncated operators.  This can be shown using the fact that $ \| \cdot \|_2 \le \| \cdot \|_1 $ for any matrix, and the orthogonality of matrix in (\ref{truncatedc}) with the discarded terms (see Supplementary Material).  The same argument can be repeated for $ C_L $ and $ \delta$, where (\ref{localcoherence}) and (\ref{quantumcorrelations2}) run over the truncated set of operators and the Frobenius norm is used instead.  Defined in this way we have truncated estimators which satisfy  $ C^{\text{tr}}_L (\rho) \le C_L (\rho) $  and $ \delta^{\text{tr}} (\rho) \le \delta (\rho) $.

%

\begin{figure}[t]
\includegraphics[width=\columnwidth]{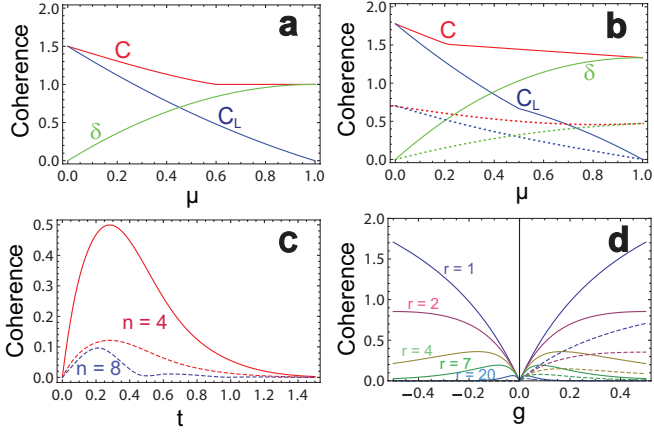} 
\caption{The coherence  $ C $ of various bipartite quantum states including their contributions from the local coherence $ C_{\text{L}}$ and global correlation $ \delta $. The states and measurement operators are: 
(a)  $| \Psi_S \rangle  = |+\rangle | + \rangle $ and $| \Psi_E \rangle = (|0\rangle | 0 \rangle + |1\rangle | 1 \rangle)/\sqrt{2} $ in (\ref{2qubitstate1}) with Pauli matrices $ {\cal A}_l,  {\cal B}_{l'} \in \{ I, \sigma^x,  \sigma^y,  \sigma^z \}/\sqrt{2} $.  
(b) $| \Psi_S \rangle  = (|0 \rangle + |1 \rangle + |2 \rangle) (|0 \rangle + |1 \rangle + |2 \rangle)/3 $, $| \Psi_E \rangle  =(|0\rangle | 0 \rangle + |1\rangle | 1 \rangle + |2 \rangle | 2 \rangle)/\sqrt{3} $   in (\ref{2qubitstate1}). 
Coherence measured with Gell-Mann matrices $ {\cal A}_l,  {\cal B}_{l'}  \in {\cal G}_{\text{SU}(3)} = \{\sqrt{2/3} I ,\lambda^{(1)}, \dots,  \lambda^{(8)} \}/\sqrt{2} $ (solid lines), truncated spin basis 
$ {\cal A}_l,  {\cal B}_{l'}  \in {\cal L}_{n=2} $ (dashed lines) where $ {\cal L}_{n} = \{ \sqrt{n(2+n)/3} I, S^x, S^y,S^z \}/\sqrt{n(1+n)(2+n)/3} $ are the orthonormal spin-$ n/2 $ operator set.  
(c) Total coherence of spin squeezed state evolved under Markovian dephasing with rate $ \Gamma = 1 $ for $n$ qubit spin ensembles.  Coherence measured with Gell-Mann matrices $ {\cal A}_l,  {\cal B}_{l'}  \in {\cal G}_{\text{SU}(n)} $ (solid lines) and truncated operators  ${\cal A}_l,  {\cal B}_{l'} \in {\cal L}_{n}  $ (dashed lines).  
(d) Total coherence of the ground state of the generalized AKLT model between two sites separated by $ r $ sites.  Coherence measured using ${\cal A}_l,  {\cal B}_{l'}  \in {\cal G}_{\text{SU}(3)} $.   For the truncated operators  $ {\cal A}_l,  {\cal B}_{l'} \in {\cal L}_{n=2}$, using estimators coherences are lower for $ g\ge 0 $ but zero for $ g \le 0 $. }  \label{fig1}  
\end{figure}

\paragraph{Example 1: Qubit and qutrit bipartite systems}

We now calculate several examples to show the theory in action.  We consider a density matrix of the form
\begin{equation}
\rho = (1-\mu) | \Psi_S \rangle \langle \Psi_S |  + \mu | \Psi_E \rangle \langle \Psi_E |,
\label{2qubitstate1}
\end{equation}
where $| \Psi_S \rangle $ and $ | \Psi_E \rangle $ are separable and entangled states.  The mixing parameter  $\mu \in [0,1] $ changes the character of the state from a separable state to an entangled state. In Fig. \ref{fig1}(a) shows the case for two qubits, with $| \Psi_S \rangle  = |+\rangle |+ \rangle $, and $| \Psi_E \rangle 
= (|0 \rangle | 0 \rangle + |1 \rangle | 1 \rangle)/\sqrt{2} $. As expected the local coherence $ C_{\text{L}} $ is a maximum for $ \mu = 0 $, and the global correlation contribution $ \delta $ is a maximum for $ \mu = 1 $. The total coherence is found to have a larger value for the $ |+ \rangle | + \rangle $ state, rather than the Bell state.  This is in agreement to expectation since in the $ \{ |0\rangle, | 1 \rangle \} $ basis, the $ |+ \rangle | + \rangle $ state has all off-diagonal elements occupied, in contrast to the Bell state which has only specific coherent elements.  

We now compare the effects of truncating the operator basis for the measurements.  In this case we consider two qutrits with  $| \Psi_S \rangle  = (|0 \rangle + |1 \rangle + |2 \rangle) (|0 \rangle + |1 \rangle + |2 \rangle)/3 $, $| \Psi_E \rangle  =(|0\rangle | 0 \rangle + |1\rangle | 1 \rangle + |2 \rangle | 2 \rangle)/\sqrt{3} $.  We compare two approaches of keeping the full SU(3) Gell-Mann basis including the identity to give $ N = 81 $ and  a truncated set  where only three SU(2) spin-1 matrices and identity are kept to give $ N = 16 $.  The two cases are shown in Fig. \ref{fig1}(b).  We see that the two cases give qualitatively the same behavior, with again the local coherence obeying a trade-off behavior with the global correlation.  From the choice of the measurement operators we can see that the overall qualitative nature of the quantum coherence does not change in the system, and the truncated expressions for the coherence are lower than the genuine values.  The quantitative difference arises due to the fact that the smaller set of operators does not capture the entire coherence in system, and the Frobenius norm is used.  It is natural to expect that a smaller set of operators does not completely capture the coherence in the system, but selectively quantifies various types of coherence.

\paragraph{Example 2: Spin squeezed state}

We also show some examples of higher dimensional systems which are promising from a practical perspective where complete tomography is difficult.  We consider a two-axis spin squeezed state between two spin ensembles, generated by the Hamiltonian
\begin{align}
H = S^-_A S^-_B + S^+_A S^+_B ,
\label{squeezedstate}
\end{align}
where $ S^{\pm}_{A,B} = \sum_{j=1}^n \sigma^{\pm}_j $ are total spin operators of an ensemble of $n $ qubits, and $ A,B $ label the two ensembles. This Hamiltonian is applied to the state $ |S^z = n \rangle |S^z = n \rangle $ in the presence of  Markovian $ S^z $-dephasing with rate $ \Gamma $ evolving with a Lindbladian master equation during the evolution of (\ref{squeezedstate}) producing a mixed state (see Supplementary Information).  In the limit of large $ n $, the Hamiltonian (\ref{squeezedstate}) is equivalent to a two-mode squeezing interaction state under the Holstein-Primakoff transformation \cite{braunstein2005quantum}.  Figure \ref{fig1}(c) shows the total coherence for the cases of $ n = 4, 8 $ with and without a full operator basis.  In this case the local coherence $ C_{\text{L}} = 0 $ for all time and $ \delta =C $ since the reduced density matrix is always completely diagonal.  The truncated operator basis is still effective at capturing the coherence in the system which is lower than the coherence using the full operator basis as expected.   Using the truncated operator basis gives a great reduction in computational overhead, and we plot the $ n = 8 $ case (along with $ n = 4$) which shows the expected reduction in timescale for reaching the maximal coherence,  scaling with $ \sim 1/\sqrt{n} $.

\paragraph{Example 3: Generalized AKLT model}

We next consider the the generalized Affleck-Kennedy-Lieb-Tasaki (AKLT) model with periodic boundary conditions \cite{affleck2004rigorous,klumper1993matrix,alipour2007entanglement}, with the Hamiltonian 
\begin{align}
&H  = \sum_{j=1}^{M} (1+2g^2)  \vec{S}_{j} \cdot \vec{S}_{j+1} + (\vec{S}_{j} \cdot \vec{S}_{j+1})^{2} + 2(1-g)\Big[  \nonumber \\
& 2(1+g) (S_{j}^z)^{2} -(1-g) (S_{j}^z S_{j+1}^z )^{2} - g \{\vec{S}_{j} \cdot \vec{S}_{j+1}, S_{j}^z S_{j+1}^z \} \Big], \nonumber 
\end{align}
where the $ \vec{S}_{j} $ are spin-1 operators ($n=2$) on site $ j $. This Hamiltonian is the AKLT model for $ g=1 $, and a critical point is present at $ g = 0 $ (see Supplementary Material). This model has a ground state that takes the form of a matrix product state, and all spin correlation can be written 
down in the thermodynamic limit exactly \cite{alipour2007entanglement}.   Using the reduced density matrix between site 1 and site $ r $, we calculate the quantum coherence using both the full SU(3) Gell-Mann basis forming a $ 9 \times 9 $ observable matrix, and the reduced SU(2) spin-1 operators forming a $ 4 \times 4 $ observable matrix (see Fig. \ref{fig1}(d)).  We find again that the local coherence is zero and  $ \delta = C  $ for all $ g $.  At $g=0$ both the full and truncated basis set exhibits non-analytic behavior which signals the presence of a quantum phase transition, for any value of $ r $. The truncated basis, using Schatten-$1$ norm, has identical results to the full basis for $ g\ge 0 $ whereas Frobenius norm gives lower amount of coherence, but is zero for $ g \le 0 $ using both norms. The reason for this can be understood as from the nature of the transition in this model where XY ferromagnetism exists for $ g> 0 $ and all XY-like spin correlators are zero for $ g \le 0 $. Since the $S_{z}$-$S_{z} $ correlations do not contribute to the coherence, our SU(2) coherence is zero for $ g \le 0 $.  Thus our reduced basis contains correlations that are common to a ferromagnetic order parameter, resulting in the same type of behavior.  
\\

\paragraph{Conclusions}
In this paper we have derived expressions to quantify the coherence and its contribution due to local coherence and global correlations in a quantum system based on expectation values of observables. 
For the case that an informationally complete measurement is made, the full coherence of the system is recovered.  For an incomplete set of measurements, the evaluated coherence forms an approximate lower bound to the value using a complete set of measurements.  The approximation only comes about due to the use of Schatten-$1$ norm which is shown to obey the full properties of a coherence.  For the Frobenius norm, the bound is exactly satisfied, although no longer obeys all the axioms of coherence, which is less desirable.  
Our observable approach allows one to directly use measurement results in a simple way to estimate the coherence in a straighfoward and efficient way.  This is especially suitable for high dimensional systems such as condensed matter physics \cite{malvezzi2016quantum,radhakrishnan2017quantum,radhakrishnan2017quantum2} and many-body atomic systems \cite{schmied2016bell,byrnes15,ilo2014theory} where complete tomography of the quantum state would be highly resource intensive.

This work is supported by the Shanghai Research Challenge Fund; New York University Global Seed Grants for Collaborative Research; National Natural Science Foundation of China (61571301,D1210036A); the NSFC Research Fund for International Young Scientists (11650110425,11850410426); NYU-ECNU Institute of Physics at NYU Shanghai; the Science and Technology Commission of Shanghai Municipality (17ZR1443600); the China Science and Technology Exchange Center (NGA-16-001); and the NSFC-RFBR Collaborative grant (81811530112).

\bibliographystyle{apsrev}
\bibliography{ref}

\appendix
\section{Appendix A: Invariance of the coherence under global unitary transformations}

In this section, we show that the definition of coherence 
\begin{align}
C (\rho) & = \| \Omega(\rho) - \Omega(\rho^{d}) \|_1
\label{coherencesuppl}
\end{align}
is invariant under unitary transformations for the Schatten-$p$ norms. 

The observables under a unitary transformation $ \rho \rightarrow U^\dagger \rho U $ transform as
\begin{align}
& \Omega_{kk'} (\rho)  = \frac{e^{i\pi/4} \text{Tr}( {\cal M}_{kk'} \rho ) 
+ e^{-i\pi/4} \text{Tr}( {\cal M}_{k' k} \rho )}{\sqrt{2}e^{i \pi \text{sgn}(k'-k)/4}}   \nonumber \\
& \rightarrow  \frac{e^{i\pi/4} \text{Tr}( {\cal M}_{kk'} U^\dagger \rho U ) 
+ e^{-i\pi/4} \text{Tr}( {\cal M}_{k' k} U^\dagger \rho U )}{\sqrt{2}e^{i \pi \text{sgn}(k'-k)/4}} \nonumber \\
& =  \frac{e^{i\pi/4} \text{Tr}(U  {\cal M}_{k k'} U^\dagger \rho ) 
+ e^{-i\pi/4} \text{Tr}( U {\cal M}_{k' k} U^\dagger \rho  )}{\sqrt{2}e^{i \pi \text{sgn}(k'-k)/4}} \nonumber \\
& = \sum_{l l'} \frac{e^{i\pi/4} O_{kl} O_{k' l'}  \langle {\cal M}_{l l'}  \rangle
 + e^{-i\pi/4} O_{kl} O_{k' l'}  \langle {\cal M}_{l' l}  \rangle }{\sqrt{2}e^{i \pi \text{sgn}(k'-k)/4}} \nonumber \\ 
& = \sum_{l l'} O_{k l} \Omega_{l l'} ( \rho) O_{l' k'}^T
\end{align}
where $O $ is an orthogonal matrix.  Covariance matrices transform in the same way under a unitary transformation \cite{gittsovich2008unifying}. The coherence is invariant under such transformations since
\begin{align}
C (U^\dagger \rho U)  & = \| O \Omega(\rho) O^T - O \Omega(\rho_{d}) O^T \|_1 \nonumber \\
 & = \| \Omega(\rho) - \Omega(\rho^{d}) \|_1= C (\rho)  ,
\end{align}
where we used the fact that the Schatten-$p$ norm is invariant under orthogonal transformations.

\section{Appendix B: Validity as a coherence measure}


\subsection{1. Coherence conditions by Yu et al.}

In this section we show that our definition of coherence (\ref{coherencesuppl}) is a valid coherence measure, under properties (C1$'$), (C2$'$), and (C3$'$) in Ref. \cite{yu2016alternative}.  In the notation of Ref.  \cite{yu2016alternative}, (C1$'$)=(C1), (C2$'$)=(C2), (C3$'$)=(C3).

\subsubsection{a. Property (C1$'$)}

Property (C1$'$) states that a coherence measure should satisfy $C(\rho) = 0 $ if and only if $ \rho  \in \cal I $, and $ C(\rho) \ge 0 $.  This property follows from the fact that the full set of observables $ \Omega(\rho) $
constitutes a complete description of the density matrix, since $\{ {\cal M}_{k k'} \}$ forms a complete operator basis. The fact that $ \Omega(\rho) $ constitutes a complete description of the density matrix can be easily verified for the explicit choice of the observables $\{ {\cal M}_{k k'} \}$. We can explicitly evaluate
\begin{align}
\langle {\cal M}_{k k'} \rangle = \left\{
\begin{array}{cc}
\rho_{k k} & \hspace{5mm} k'=k  \\
\sqrt{2} \text{Re} ( \rho_{k k'} ) & \hspace{5mm} k'> k \\
\sqrt{2} \text{Im} ( \rho_{k k'} ) & \hspace{5mm} k'< k 
\end{array}
\right.  
\label{specificms}
\end{align}
where $ \rho_{k k'} = \langle k | \rho | k' \rangle $.  Substitution into Eq. (1) of the main text then gives 
\begin{align}
\Omega (\rho) = \rho .
\label{omegamatrixelem}
\end{align}
For perfect measurements the observable matrix is the density matrix itself, hence contains all information of the quantum state. 

The property (C1$'$) follows directly from the property of any matrix norm that $ \| A \| = 0  $ if and only if $ A = 0 $.  Using (\ref{omegamatrixelem}) with (\ref{coherencesuppl}),  the coherence is zero if and only if $ \rho = \rho^d $.  Since $ \rho^d $ constitutes the set of incoherent states $ \cal I $, this shows $C(\rho) = 0 $ if and only if $ \rho \in \cal I $.  The fact that $ C(\rho) \ge 0 $ follows from the properties of any matrix norm.

\subsubsection{b. Property (C2$'$)}

Property (C2$'$) states that under an incoherent completely positive and trace preserving (ICPTP) operation $ \Lambda $, the coherence is non-increasing $ C(\rho) \ge C( \Lambda (\rho)) $.  This is identical to property (C2a) in Ref. \cite{baumgratz2014quantifying}. First note that for perfect measurements we have (\ref{omegamatrixelem}), and hence under the ICPTP map the coherence is 
\begin{align}
C(\Lambda(\rho)) = \| \Lambda(\rho) - \rho^d \|_1  .
\label{icptpcoh}
\end{align}
In Ref. \cite{perez2006contractivity} it has been shown that the Schatten-$1$ norm is contracting under CPTP maps.  Since (\ref{icptpcoh}) follows this form, (C2$'$) is satisfied.

\subsubsection{c. Property (C3$'$)}

Property (C3$'$) states that a coherence measure must satisfy
\begin{align}
C(p_1 \rho_1 \oplus p_2 \rho_2 )  =p_1 C(\rho_1) + p_2 C(\rho_2)
\label{yuc3}
\end{align}
where $ \rho_1 $ and $ \rho_2 $ are density matrices in different subspaces and $ p_1 + p_2 = 1 $ are probabilities. Writing these matrices explicitly, we have
\begin{align}
\rho_1 & = 
\left(
\begin{array}{cc}
\sigma_1 & 0 \\
0 & 0 
\end{array}
\right)  \nonumber \\
 \rho_2 & = 
\left(
\begin{array}{cc}
0 & 0 \\
0 & \sigma_2
\end{array}
\right)  \nonumber \\
\rho & = p_1 \rho_1 \oplus p_2 \rho_2 = \left(
\begin{array}{cc}
p_1 \sigma_1  & 0 \\
0 & p_2 \sigma_1 
\end{array}
\right) .
\end{align}
where $ \sigma_1 , \sigma_1  $ are the submatrices for the density matrices in the two spaces, and are density matrices themselves.  

The matrix elements of the observable matrix for each of these states are also block diagonal, using the expression (\ref{omegamatrixelem}).  
\begin{align}
\Omega(\rho_1) & = \left(
\begin{array}{cc}
\Omega(\sigma_1) & 0 \\
0 & 0 
\end{array}
\right)  \nonumber \\
\Omega(\rho_2) & = \left(
\begin{array}{cc}
0& 0 \\
0 & \Omega(\sigma_2)  
\end{array}
\right)  \nonumber \\
\Omega(\rho) & = \left(
\begin{array}{cc}
p_1 \Omega(\sigma_1) & 0 \\
0 & p_2 \Omega(\sigma_2)
\end{array}
\right) 
\end{align}
Here we have taken the measurement basis such that the subspace structure is preserved, i.e. any basis that admixes states within the same subspace.  

Evaluating the coherence according to (\ref{coherencesuppl}), we have for the right hand side of (\ref{yuc3})
\begin{align}
C(\rho_1) & = \| \Omega(\sigma_1) - \Omega(\sigma_1^d) \|_1 \nonumber \\
C(\rho_2) & = \| \Omega(\sigma_2) - \Omega(\sigma_2^d) \|_1 
\label{rhsyu}
\end{align}
since the matrix norm only depends upon the non-zero submatrices.  Here we have defined the dephased submatrices $ \sigma_1^d,  \sigma_2^d $ which only contain the diagonal components of  $ \sigma_1,  \sigma_2 $ respectively.  The left hand side of (\ref{yuc3}) is 
\begin{align}
C(\rho) &  = \| p_1 \Omega(\sigma_1) - p_1 \Omega(\sigma_1^d) \|_1 + \| p_2 \Omega(\sigma_2) - p_2 \Omega(\sigma_2^d) \|_1 \nonumber \\
& = p_1 \|  \Omega(\sigma_1) -  \Omega(\sigma_1^d) \|_1 + p_2  \| \Omega(\sigma_2) -  \Omega(\sigma_2^d) \|_1 .
\label{lhsyu}
\end{align}
In the first line we used the fact that the trace norm of a block diagonal matrix is the sum of the trace norms of the submatrices. In the second line we used the absolutely homogenous property of matrix norms. Substitution of (\ref{rhsyu}) and (\ref{lhsyu}) into (\ref{yuc3}) verifies property (C3$'$).

\subsection{2. Coherence conditions by Baumgratz et al.}

In this section we show that our definition of coherence (\ref{coherencesuppl}) is a valid coherence measure, under properties (C1), (C2b), and (C3) in Ref. \cite{baumgratz2014quantifying}.

\subsubsection{a. Property (C1)}

Property (C1) in Ref. \cite{baumgratz2014quantifying} is identical to property (C1$'$) in Ref. \cite{yu2016alternative}.  This is verified from the same arguments as Sec. $1.a$ .

\subsubsection{b. Property (C2b)}

Property (C2b) states that the average coherence should decrease after performing an ICPTP map
\begin{align}
C(\rho) \ge \sum_n q_n  C(\sigma_n) ,
\label{propertyc2b}
\end{align}
where the state after measurement is
\begin{align}
\sigma_n = \frac{1}{q_n} K_n \rho K_n^\dagger ,
\end{align}
and the probability of this outcome is
\begin{align}
q_n = \text{Tr} (K_n \rho K_n^\dagger ),
\end{align}
where $K_n $ is the Kraus operator for an ICPTP map. 

Consider the basis $ \{ |l \rangle \} $ to be the basis that defines the incoherent states, such that a general decohered state is written
\begin{align}
\rho^d = \sum_{l} \rho_{ll} | l \rangle \langle l | .
\end{align}
The general form of the Kraus operator for an ICPTP map can be written
\begin{align}
K_n = \sum_l c_l^n |P_n(l) \rangle \langle l | ,
\label{expliciticptp}
\end{align}
where $ P_n(l) $ is a permutation function which permutes the $ l \in [1,D ] $ labels. The $ c_l
^n $ are complex coefficients.  To satisfy the requirement that  $ \sum_n K_n^\dagger K_n = I $, the coefficients must satisfy 
\begin{align}
\sum_n | c_l^n |^2 = 1 .
\label{c2b_cnl}
\end{align}
The structure of (\ref{expliciticptp}) ensures that it is an ICPTP map. Applying it to the state $ \rho^d $, we obtain
\begin{align}
K_n \rho^d K_n^\dagger = \sum_l |c_l^n|^2
 \rho_{ll} | P_n(l)
\rangle \langle P_n(l) | , \label{c2b_KrhoKd}
\end{align}
which is another diagonal state and is contained in $ \cal I $.

Starting with the right hand side of (\ref{propertyc2b}), we have in our case
\begin{align}
 \sum_n q_n  C(\sigma_n) = \sum_n q_n \| \Omega(\sigma_n) - \Omega(\sigma_n^d) \|_1 .
\label{funeq}
\end{align}
Substituting the definitions it follows that
\begin{align}
\Omega_{kk'} (\sigma_n) & = 
\frac{1}{q_n} \sum_{l l'} \rho_{l l'} c_l^n {c_{l'}^n}^*  \langle P_n(l') |k \rangle \langle k' | P_n(l) \rangle \nonumber \\
& = \frac{1}{q_n} 
 \rho_{P_n^{-1}(k') P_n^{-1}(k)} c_{P_n^{-1}(k')}^n (c_{P_n^{-1}(k)}^n)^* ,
\label{omegakexp}
\end{align}
where we used the fact that $ \langle P_n (l) | P_n (l') \rangle = \delta_{l l'} $ and 
\begin{align}
q_n = \sum_l |c_l^n|^2 \rho_{ll} .
\end{align}
Meanwhile, the state $ \sigma_n^d $ is the decohered counterpart of $ \sigma_n $ which is
\begin{align}
\sigma_n^d  & = \sum_l |l \rangle \langle l | \langle l | \sigma_n | l \rangle \nonumber \\
& = \frac{1}{q_n} \sum_l |c_l^n|^2 \rho_{ll} | P_n(l) \rangle \langle P_n(l) | .
\end{align}
Evaluating the observable matrix for the decohered state we thus obtain
\begin{align}
\Omega_{kk'} (\sigma_n^d) & = \frac{ \delta_{k k'}}{q_n} 
\rho_{P_n^{-1}(k) P_n^{-1}(k)} | c_{P_n^{-1}(k)}^n |^2 ,
\end{align}
which takes the same form as (\ref{omegakexp}) but with just diagonal terms in the expectation value.  Thus the difference can be written as
\begin{align}
& q_n (\Omega_{kk'} (\sigma_n) - \Omega_{kk'} (\sigma_n^d) ) = \nonumber \\
& (1- \delta_{kk'}) \rho_{P_n^{-1}(k') P_n^{-1}(k)} c_{P_n^{-1}(k')}^n (c_{P_n^{-1}(k)}^n)^*   .
\label{observablec2b}
\end{align}

Let us now define the matrices with elements
\begin{align}
\rho^{(n)}_{k k'} &= c^{n}_k (c^{n}_{k'})^* \rho_{kk'} \\
R^{(n)}_{k k'} & =  (1- \delta_{kk'}) \rho^{(n)}_{k k'} 
\label{rndef}
\end{align}
Substituting (\ref{observablec2b}) into the matrix norm according to (\ref{observablec2b}) we obtain
\begin{align}
 \sum_n q_n  C(\sigma_n) =  \sum_n \| R^{(n)} \|_1  . 
\end{align}
Here we have used the fact that for a matrix norm, the ordering of the matrix elements is arbitrary and gives the same value under a permutation $ k \rightarrow P_n(k) $.  Using similar steps we find that the left hand side of (\ref{propertyc2b}) can be evaluated to be 
\begin{align}
C(\rho) =  \| R \|_1 ,
\label{standardcoh}
\end{align}
where the matrix $ R $ has elements
\begin{align}
R_{k k'} & =  (1- \delta_{kk'}) \rho_{k k'} .
\label{rdef}
\end{align}
Thus showing property (C2b) amounts to showing that 
\begin{align}
 \| R \|_1 \ge  \sum_n \| R^{(n)} \|_1 .\label{c2b}
\end{align}
We now show that this is obeyed for several cases as shown below.

\subsubsection{c. Proof of (C2b) property for X-states}

In order to prove the correctness of (\ref{c2b}) for the Schatten-$1$ norm we need to write it explicitly in terms of eigenvalues of $R^{(n)}$ and $R$. For X-states the density matrices have a very convenient form consisting exclusively of anti-diagonal, making it possible to define eigenvalues for any number of qubits. For system of $q$ qubits assuming $q$ is an even number we have
\begin{align}
\vert \lambda^{(n)}_{l} \vert = \vert \lambda^{(n)}_{q - l} \vert = \sqrt{R^{(n)}_{l,q-l} R^{(n)}_{q-l,l}} .\label{defEigen}
\end{align}
In case of odd $q$ there is an extra eigenvalue equal to the element located at 
$R^{(n)}_{\text{center}}$ which is the center element of matrix $R^{(n)}$. We can ignore this case, since from definition (\ref{rndef}) we know that the diagonal elements are all $0$'s, and thus for odd dimensional $n\times n$ matrix, center element $\big((n+1)/2, (n+1)/2\big)$ eventually becomes $0$.

As before, we have a set of $n$-independent eigenvalues $\lambda_l$ which are dependent on $R$ instead of $R^{(n)}$. In the definition of Schatten-$1$ norm under the absolute value half of those eigenvalues become degenerate creating a constant factor on both sides of (\ref{c2b}), thus we can use first half of them and rewrite (\ref{c2b}) for X-states as
\begin{align}
\sum_n \sum_{l \leq \frac{q}{2}} \vert \lambda^{(n)}_{l} \vert \leq \sum_{l \leq \frac{q}{2}} \vert \lambda_{l} \vert .
\end{align}
Substituting from definition (\ref{defEigen})
\begin{align}
\sum_n \sum_{l \leq \frac{q}{2}} \vert \sqrt{R^{(n)}_{l,q-l} R^{(n)}_{q-l,l}} \vert \leq \sum_{l \leq \frac{q}{2}} \vert \sqrt{R_{l,q-l} R_{q-l,l}} \vert . \label{c2b_p0}
\end{align}
We can evaluate from the definitions that 
\begin{align}
\sqrt{R_{l,q-l} R_{q-l,l}} &= \delta_{l, q-l} \vert \rho_{q-l,l} \vert \\
\sqrt{R^{(n)}_{l,q-l} R^{(n)}_{q-l,l}} &= \delta_{l, q-l} \vert \rho^{(n)}_{q-l,l} \vert \\
&= \delta_{l, q-l} \vert c^{(n)}_{q-l} c^{*(n)}_{l} \rho_{q-l,l} \vert .
\end{align}
%
%
%
Continuing from (\ref{c2b_p0}), we use the Idempotence property to simplify nested absolute values, leading us to simpler form
\begin{align}
\sum_n \sum_{l \leq \frac{q}{2}} \vert c^{(n)}_{q-l} c^{*(n)}_{l} \rho_{q-l,l} \vert \leq \sum_{l \leq \frac{q}{2}} \vert\rho_{q-l,l} \vert .
\end{align}
We can take advantage of the fact that $\vert\rho_{q-l,l} \vert$ has no $n$-dependence to rearrange the sums on the left-hand-side
\begin{align}
\sum_{l \leq \frac{q}{2}} \vert \rho_{q-l,l} \vert ( \sum_n \vert c^{(n)}_{q-l} c^{*(n)}_{l} \vert ) \leq \sum_{l \leq \frac{q}{2}} \vert\rho_{q-l,l} \vert  . \label{c2b_p3}
\end{align}
At this point we can see that (\ref{c2b_p3}) is true as long as coefficients 
\begin{align}
\sum_n \vert c^{(n)}_{q-l} c^{*(n)}_{l} \vert \leq 1  .\label{c2b_condition}
\end{align}
Applying Cauchy-Schwarz inequality to (\ref{c2b_condition}) we find
\begin{align}
\sum_n \vert c^{(n)}_{q-l} c^{(n)}_{l} \vert \leq (\sum_n \vert c^{(n)}_{q-l} \vert^2)^\frac{1}{2} (\sum_n \vert c^{(n)}_{l} \vert^2 )^\frac{1}{2} .\label{c2b_cauchy_schwarz}
\end{align}
From (\ref{c2b_cnl}) we can see that both factors in the product on the right hand side of (\ref{c2b_cauchy_schwarz}) are equal to $1$ and thus (\ref{c2b_condition}) is verified which proves (\ref{c2b_p3}) and it implies 
(\ref{propertyc2b}) for $ X $-states.

\subsubsection{d. Numerical verification of (C2b) property for general $\rho$}

For $\rho$ that is not of the form of an $X$-state form we run numerical tests to verify (\ref{c2b}).
A single test instance consists of matrix $R$ and set of matrices $R^{(n)}$ constructed according to the definitions (\ref{rndef}) and (\ref{rdef}). For each such test instance we used the QuTip library in Python to generate random density matrices $\rho$ and sets of random kets for the complex coefficients $c^{(n)}_l$.
We tried a total of 1 million random instances for the parameters as shown in  Table \ref{tabc2b} and found no violations.

\begin{table}[t]
\begin{center}
\begin{tabular}{ |c|c| } 
 \hline
dimension $D$ & number of $K_n$  \\ 
\hline
2 & 2 \\ 
2 & 4 \\ 
4 & 2  \\ 
4 & 4 \\ 
8 & 2  \\ 
8 & 4  \\ 
16 & 2 \\ 
16 & 4  \\ 
32 & 2 \\ 
32 & 4  \\ 
 \hline
\end{tabular}
\caption{Parameters chosen for the numerical verification of the inequality (\ref{c2b}) for the Schatten-$1$ norm.  For each parameter set, $10^5 $ instances were randomly generated and no violations were found.  \label{tabc2b}}
\end{center}
\end{table}

\subsubsection{\label{sec:level1}e. Example: Satisfaction of (C2b) for a specific ICPTP map}

We show that (C2b) is satisfied for the example as that given in Sec. VII of the Supplementary Material in Ref. \cite{baumgratz2014quantifying} for the Schatten-$1$ norm, which was a counterexample for the Frobenius norm. The Kraus operators are taken to be
\begin{align}
K_1 = \left(
\begin{array}{ccc}
0 & 1 & 0 \\
0 & 0 & 0 \\
0 & 0 & \alpha 
\end{array}
\right), \nonumber \\
K_2 = \left(
\begin{array}{ccc}
1 & 0 & 0 \\
0 & 0 & \beta \\
0 & 0 & 0 
\end{array}
\right) ,
\end{align}
where $\alpha,\beta \in \mathbb{C} $ and $|\alpha|^2 + |\beta|^2 =1 $. 
The state that we take is 
\begin{align}
\rho = \mu | \psi_1 \rangle \langle \psi_1 | + (1-\mu) | \psi_2 \rangle \langle \psi_2 |,
\end{align}
where $ | \psi_1 \rangle = [0 1 0 ]^T $ and $ | \psi_2 \rangle = [1 0 1 ]^T/\sqrt{2} $. We find $ C(\rho) = 1-\mu $ and  $\sum_n q_n  C(\sigma_n) = (1-\mu)|\beta| $. This satisfies (C2b) for all $ \mu, \alpha, \beta $.

\subsubsection{f. Property (C3)}

Property (C3) states that the coherence should decrease with mixing of quantum states, in comparison to the average of the original coherences
\begin{align}
C(\sum_n p_n \rho_n ) \le \sum_n p_n C(\rho_n) ,
\label{propertyc3}
\end{align}
where $ p_n $ are arbitrary mixing probabilities and $ \rho_n $ are an arbitrary set of density matrices. 

Starting from the left hand side of (\ref{propertyc3}), we have in our case
\begin{align}
C(\sum_n p_n \rho_n ) = \| \Omega(\sum_n p_n \rho_n) - \Omega(\sum_n p_n \rho_n^d) \|_1 .
\end{align}
The observables are linear in the expectation values hence 
\begin{align}
C(\sum_n p_n \rho_n ) & = \| \sum_n p_n ( \Omega(\rho_n) -\Omega(\rho_n^d) ) \|_1 \nonumber \\ 
& \le \sum_n p_n  \|\Omega(\rho_n) -\Omega(\rho_n^d)  \|_1 \nonumber \\ 
& = \sum_n p_n C(\rho_n) ,
\end{align}
where we used the the subadditivity of the matrix norm.

\subsection{3. Violation of coherence property for the covariance matrix}

The global correlation $ \delta(\rho) $ as defined in the main text in general does not satisfy the properties for a coherence.  An exception to this is when the global correlation has the full contribution to the total coherence. Here we show an explicit example of a violation of property (C3).

 We consider a similar example as given in Fig. 1(a) of the main text for the global correlation $\delta(\rho)$, under mixing of the states $ |\Psi_S \rangle $ and $ | \Psi_E \rangle $ with mixing parameter $\mu $ . We evaluate $ C(\sum_n p_n \rho_n) = 1-(1-\mu)^2 $ and $ \sum_n p_n C(\rho_n) = \mu $. Here, the property (C3) is violated at any value of $\mu \in (0,1)$. We attribute the failure of $ \delta (\rho) $ as a coherence measure due to the quadratic term in the definition of the covariance matrix. The quadratic probability terms are not canceled when the coherences are averaged. We note that property (C3) is also found to be violated for the covariance matrix when using the $l_1$-norm.

\section{Appendix C: Incomplete sets of operators}

\subsection{\label{sec:level1}1. Estimator based on Frobenius norm}

In this section we show that we may construct an estimator for the coherence using a truncated set of operators that is upper bounded by the genuine value of the coherence.  Consider the expression for the coherence given in Eq. (5) of the main text. The full expression for the coherence can be written
\begin{align}
C(\rho) = \| P + Q \|_1 ,
\end{align}
where
\begin{align}
P & = \sum_{l \in {\cal O}_{\text{tr}}} {\cal S}_l ( \langle {\cal S}_l \rangle_{\rho} - \langle {\cal S}_l \rangle_{\rho^d} ) \\
Q & =  \sum_{l \in \bar{{\cal O}}_{\text{tr}}} {\cal S}_l ( \langle {\cal S}_l \rangle_{\rho} - \langle {\cal S}_l \rangle_{\rho^d} ) 
\end{align}
and $ {\cal O}_{\text{tr}} $ are the set of terms that are kept in the truncation and $ \bar{{\cal O}}_{\text{tr}} $ are the remaining terms.  The coherence using the truncated operator set is defined using the Frobenius norm, as discussed in the main text
\begin{align}
C_{\text{tr}}(\rho) = \| P \|_2 .
\end{align}

We wish to show that $ C_{\text{tr}}(\rho) \le C(\rho) $, or equivalently, 
\begin{align}
\| P \|_2 \le \| P + Q \|_1.
\label{pqrel}
\end{align}
First note that from the general properties of matrix norms
\begin{align}
\| P + Q \|_2  \le \| P + Q \|_1  .
\label{frobschatt}
\end{align}
The left hand side can be simplified in our case since
\begin{align}
\| P + Q \|_2 & =  \sqrt{  \text{Tr} (P^\dagger + Q^\dagger)(P+Q) } \nonumber \\
& = \sqrt{ \text{Tr} (  P^\dagger P + P^\dagger Q + Q^\dagger P + Q^\dagger Q)} \nonumber \\
& =  \sqrt{  \text{Tr} ( P^\dagger P  +  Q^\dagger Q) }
\end{align}
where we used the fact that the set of operators $ \{ {\cal S}_l \} $ are orthonormal, such that 
\begin{align}
\text{Tr} ( P^\dagger Q ) = \text{Tr} ( Q^\dagger P ) = 0 .  
\label{linind}
\end{align}
Then 
\begin{align}
 \| P \|_2^2 = \text{Tr} ( P^\dagger P)  \le  \text{Tr} (P^\dagger P  +  Q^\dagger Q ) =  \| P + Q \|_2^2
\label{frobrel}
\end{align}
since $ Q^\dagger Q $ is a positive semidefinite operator. Combining (\ref{frobschatt}) and (\ref{frobrel}) shows (\ref{pqrel}).

\subsection{2. Estimator based on Schatten-1 norm}

A simpler procedure for the truncated estimator would be
\begin{align}
\| P \|_1 \le \| P + Q \|_1 \hspace{1cm} \text{(false)}
\label{pqrelorig}
\end{align}
Unfortunately, for the Schatten-$1$ norm, (\ref{pqrelorig}) does not always hold.  We have numerically generated random matrices to test the frequency and level of violation of the inequality  (\ref{pqrelorig}), the results are shown in Table \ref{tab1}.  Violations are found for any matrix dimension larger than 2.  However, we note that the violation of the inequality occurs infrequently and the level of violation is typically relatively small at the level $ \sim 2\% $.  Thus although not exact, the inequality  (\ref{pqrelorig}) is observed to a good approximation.  

We also note that using $ \| P \|_1 $ as the truncated estimator never falsely gives a non-zero value of coherence.  To show this, observe that 
\begin{align}
C(\rho) = \| P+ Q \|_1 = 0 
\end{align}
either if (i) $ P = - Q $, $ P \ne 0 $, and $ Q \ne 0 $; or (ii) $ P = Q = 0 $.  However, since $ P $ and $ Q $ are orthogonal matrices $ \text{Tr} (PQ) = 0 $ by construction, case (i) is impossible.  Hence the only time that the coherence is zero is case (ii), where $ P = 0 $.  If $ P = 0 $, then $ \| P \|_1 = 0 $, and the truncated estimator gives zero coherence.  

Thus in applications where only an estimate of the coherence is required, the truncation based on the Schatten-$1$ norm may be used. 
The same argument can be made for the other coherence quantities $ \delta, C_{\text{L}} $.

\begin{table}[t]
\begin{center}
\begin{tabular}{ |c|c|c| } 
 \hline
Matrix dimension & Frequency of violation & Mean violation \\ 
\hline
2 & 0 \% &  0\% \\ 
3 & 2.2\% & 3.6\% \\ 
4 & 0.3\% & 1.7\% \\ 
5 & 0.05\% & 1.8\% \\ 
 \hline
\end{tabular}
\caption{Violation frequency and level of violation of the inequality (\ref{pqrel}) for the Schatten-$1$ norm. The mean violation is the average of the quantity $ \| P \|_1 / \| P + Q \|_1 -1 $ over the violated instances. \label{tab1}}
\end{center}
\end{table}

\section{Appendix D: Spin squeezed state}

The dephased spin squeezed state is calculated using the master equation
\begin{align}
\label{lasingmaster}
\frac{d \rho}{dt} = -\frac{i}{\hbar} [ H, \rho ] -\frac{\Gamma}{2} {\cal L} [S^z,\rho] 
\end{align}
where
\begin{align}
H = S^+_A S^+_B + S^-_A S^-_B 
\end{align}
is two-axis spin squeezing Hamiltonian between two ensembles, and 
\begin{align}
{\cal L} [O,\rho] \equiv  \rho O^\dagger O + O^\dagger O \rho -  2 O \rho O^\dagger 
\end{align}
is the Lindblad superoperator. The spin operators are $ S^z = \sum_{j=1}^n \sigma^z_j, S^{\pm} = \sum_{j=1}^n \sigma^{\pm}_j $. The initial state is the completely $ S^z $ polarized state 
\begin{align}
|S^z = n \rangle |S^z = n \rangle .
\end{align}
Since the master equation is symmetric under particle interchange, we can work in the symmetric subspace which reduces the Hilbert space dimension from $ 4^n $ to $ (n+1)^2 $.

\section{Appendix E: Coherence in the generalized AKLT chain}

We calculate the coherence of the generalized Affleck-Kennedy-Lieb-Tasaki chain \cite{affleck2004rigorous,klumper1993matrix} as defined 
in Eq. (13) of the main text.  This corresponds to the Hamiltonian as given in Eq. (14) of Ref. \cite{alipour2007entanglement} with the parameters $ a=b=c=1 $ and $ \sigma = - 1 $.  

The reduced density matrix of two spins located at sites $1$ and $r$ is  \cite{alipour2007entanglement}
\begin{equation}
\rho(1,r) = 
\begin{pmatrix}
\alpha & 0 & 0 & 0 & 0 & 0 & 0 & 0 & 0 \\
0 & |g| \gamma & 0 & \mu & 0 & 0 & 0 & 0 & 0 \\
0 & 0 & \beta & 0 & \delta & 0 & 0 & 0 & 0 \\
0 & \mu & 0 & |g| \gamma & 0 & 0 & 0 & 0 & 0 \\
0 & 0 & \delta & 0 & \gamma & 0 & \delta & 0 & 0 \\
0 & 0 & 0 & 0 & 0 & |g| \gamma & 0 & \mu & 0 \\
0 & 0 & 0 & 0 & \delta & 0 & \beta & 0 & 0 \\
0 & 0 & 0 & 0 & 0 & \mu & 0 & |g| \gamma & 0 \\
0 & 0 & 0 & 0 & 0 & 0 & 0 & 0 & \alpha \\
\end{pmatrix},
\label{densitymatrixs1}
\end{equation}
where the parameters in the matrix are 
\begin{align}
\alpha & = \frac{g^{2} (\lambda_{1}^{r-2} - \lambda_{2}^{r-2})}{\lambda_{1}^{r}}, \nonumber \\
\beta & = \frac{g^{2} (\lambda_{1}^{r-2} + \lambda_{2}^{r-2})}{\lambda_{1}^{r}}, \nonumber \\
\gamma & = \frac{1}{\Lambda_{1}^{2}},\nonumber \\ 
\delta & = -g \left( \frac{g}{\Lambda_{1}} \right)^{r},\nonumber \\ 
\mu & = - |g| \left( \frac{g}{\Lambda_{1}} \right)^{r}, \nonumber \\
\lambda_{1} & = 1+2g, \nonumber \\
\lambda_{2} & = 1-2g, \nonumber \\
\Lambda_{1} & = 1+ 2|g|.
\end{align}
The decohered density matrix $ \rho^d $ is obtained by setting all off-diagonal terms to zero.  

The full SU(3) Gell-Mann operator basis corresponds to
\begin{align}
{\cal A}_0 & = \sqrt{1/3}I = \sqrt{1/3}(|0 \rangle \langle 0 | + |1 \rangle \langle 1 | + |2 \rangle \langle 2 |),  \nonumber \\
{\cal A}_1 & = \sqrt{1/2} \lambda_1 = \sqrt{3/2}( |0 \rangle \langle 1 | + |1 \rangle \langle 0 | ), \nonumber \\
{\cal A}_2 & =  \sqrt{1/2} \lambda_2 = \sqrt{3/2}( -i |0 \rangle \langle 1 | +i |1 \rangle \langle 0 | ), \nonumber \\
{\cal A}_3 & = \sqrt{1/2} \lambda_3 = \sqrt{3/2}(  |0 \rangle \langle 0 | - |1 \rangle \langle 1 | ), \nonumber \\
{\cal A}_4 & =  \sqrt{1/2} \lambda_4 =  \sqrt{3/2}( |0 \rangle \langle 2 | + |2 \rangle \langle 0 | ), \nonumber \\
{\cal A}_5 & =  \sqrt{1/2}\lambda_5 =  \sqrt{3/2}( -i |0 \rangle \langle 2 | + i |2 \rangle \langle 0 | ), \nonumber \\
{\cal A}_6 & =  \sqrt{1/2}\lambda_6 =  \sqrt{3/2}( |1 \rangle \langle 2 | + |1 \rangle \langle 2 | ), \nonumber \\
{\cal A}_7 & =  \sqrt{1/2}\lambda_7 =  \sqrt{3/2}( -i |1 \rangle \langle 2 | + i |1 \rangle \langle 2 | ), \nonumber \\
{\cal A}_8 & =  \sqrt{1/2} \lambda_8 =  \sqrt{1/2}(  |0 \rangle \langle 0 |+ |1 \rangle \langle 1 | -2 |2 \rangle \langle 2 | ) , 
\end{align}
and similarly for $ {\cal B}_l $.  These are orthonormal operators $ \text{Tr}({\cal A}_k {\cal A}_l ) =  \delta_{kl} $.  Calculating the quantum coherence using the full SU(3) Gell-Mann operator basis we obtain the results 
\begin{align}
C(\rho) & = 2(2+ \sqrt{2}) \left|  \frac{g}{(1+2 |g|)^r} \right| , \nonumber \\
C_L(\rho) & = 0 , \nonumber \\
\delta(\rho) & = C(\rho)  .
\end{align}
For the SU(2) spin-1 operator basis, we use the observables
\begin{align}
{\cal A}_0 & =\sqrt{1/3} I, \nonumber \\
{\cal A}_1 & = \sqrt{1/8}S_x^j = (\lambda_1 + \lambda_6)/2, \nonumber \\
{\cal A}_2 & = \sqrt{1/8}S_y^j = (\lambda_2 + \lambda_7)/2, \nonumber \\
{\cal A}_3 & = \sqrt{1/8}S_z^j = -\sqrt{1/8}(\lambda_3  + \sqrt{3}\lambda_8)  .
\end{align}

Calculating the quantum coherence using the SU(2) spin-1 operator basis with estimator we obtain the results 

\begin{align}
C(\rho) & =  \sqrt{2} \left|  \frac{g+|g|}{(1+2 |g|)^r} \right|, \nonumber \\
C_L(\rho) & = 0, \nonumber \\
\delta(\rho) & = C(\rho)  .
\end{align}

For the truncated SU(2) spin-1 operator basis, $ g\ge 0 $ gives precisely the same result as the full Gell-Mann operator basis for Schatten-$1$ norm, and lower amount of coherence using Frobenius norm whereas for $ g\le 0 $ the coherence collapses to zero in both cases.

\end{document}